\documentclass[a4paper,11pt]{article}
\usepackage{jcappub} 
\usepackage{lineno}
\usepackage{longtable}
\usepackage{array}

\title{\boldmath Constraints on the baryon density from fast radio bursts using a non-parametric reconstruction of the Hubble parameter}

\author{$^{1,2}$L\'azaro L. Sales}
\affiliation{$^{1}$Departamento de F\'{\i}sica, Universidade do Estado do Rio Grande do Norte,\\
59610-210, Mossor\'o-RN, Brazil}
\author{$^{2}$Klecio E. L. de Farias}
\affiliation{$^{2}$Departamento de F\'{\i}sica, Universidade Federal de Campina Grande,\\
Caixa Postal 10071, 58429-900, Campina Grande, Para\'{\i}ba, Brazil.}

\author{$^{2}$Amilcar R. Queiroz}

\author{$^{2}$Rafael A. Batista}

\author{$^{2}$Bruno W. Ribeiro}

\author{$^{2}$Raiff H. Santos}

\emailAdd{lazarolima@uern.br}

\abstract{In this study, we use a sample of 130 well-localized fast radio bursts (FRBs)
to constrain the physical baryon density $\Omega_{\rm b}h^2$, and the astrophysical contribution from host galaxies. The cosmological dependence entering the intergalactic dispersion measure is described through a non-parametric reconstruction of the Hubble parameter $H(z)$ obtained from cosmic chronometer data using the \texttt{ReFANN} neural-network framework, independently of the FRB sample. Within a Bayesian analysis, we jointly infer $\Omega_{\rm b}h^2$ and the
parameters of a log-normal host-galaxy distribution, namely its median
$e^\mu$ and logarithmic scatter $\sigma_{\rm host}$, using both real FRB data
and a mock catalog. For the real sample, we obtain
$\Omega_{\rm b}h^2=0.02236\pm0.00090$,
$e^\mu=178.15^{+16.51}_{-16.97}~\mathrm{pc}\,\mathrm{cm}^{-3}$, and
$\sigma_{\rm host}=0.794^{+0.064}_{-0.067}$. For the mock catalog, we find
$\Omega_{\rm b}h^2=0.02248\pm0.00018$,
$e^\mu=182.36^{+6.83}_{-6.48}~\mathrm{pc}\,\mathrm{cm}^{-3}$, and
$\sigma_{\rm host}=0.711^{+0.024}_{-0.025}$. The baryon density constraint
from the real FRB sample is in excellent agreement with both Big Bang
Nucleosynthesis and Planck CMB determinations, differing from their central
values by only $\simeq 0.05\%$. The mock analysis further illustrates the
potential of future FRB samples, reducing the uncertainty on
$\Omega_{\rm b}h^2$ to the sub-percent level while remaining statistically
consistent with early-Universe constraints. Our findings show that combining FRB dispersion measures with a non-parametric reconstruction of the expansion history provides a robust
pathway to constrain both cosmological and astrophysical parameters, establishing FRBs as a complementary low-redshift probe of the baryon density.}

\begin{document}
\maketitle
\flushbottom

\section{Introduction}

Fast radio bursts (FRBs) are bright, millisecond-duration radio transients of extragalactic origin. The first FRB was discovered in archival data from the Parkes radio telescope by Lorimer et al. \cite{Lorimer:2007qn}, while subsequent reanalyses of Parkes observations have identified a larger sample of events \cite{Yang:2021vfc}. They display large dispersion measures, rapid variability, and coherent emission across wide radio frequencies. At this moment, more than 4000 FRBs have been detected, where the largest part of the detections comes from the CHIME radiotelescope collaboration \cite{TheCHIMEFRB:2026nji}, where rencently has announced a total of 4539 FRBs, divided into 3641 unique sources and 83 known repeating sources producing 981 bursts. However, the localized sample with a defined redshift remains very small compared to the detections. 

The main property of FRBs lies in their dispersion measure (DM), caused by a time delay of the FRB signal across different radio frequency channels. This delay occurs due to the interactions with free electrons along the signal's path from the originating source to the detection apparatus, causing lower-frequency wave fronts to arrive later than higher-frequency ones, and from this time delay, the distance can be estimated. Because of the extremely high dispersion that surpasses the maximum contribution from the Milky Way, the FRBs are classified as extragalactic events and may be divided into repeaters (periodic or not) and non-repeaters, which correspond to a single detected pulse. Despite numerous efforts to identify their origin, the sources of these phenomena remain unclear, and the repeating signals are attributed to neutron stars with strong magnetic fields, such as pulsars \cite{Cordes:2015fua} and magnetars \cite{Popov:2007uv,Pen:2015ema}.

From a cosmological perspective, the observed DM can be decomposed into contributions from the Milky Way, the intergalactic medium (IGM), and the host galaxy, with the IGM component being directly sensitive to the baryon content of the Universe. This establishes a direct connection between FRB observations
and the baryon density parameter $\Omega_b h^2$, making FRBs useful probes of
cosmological parameters (see, e.g., Refs. \cite{Macquart:2020lln,Connor:2024mjg,wang2025probing,wu2016constraints}). However, the interpretation of FRB data depends on astrophysical uncertainties associated with the host galaxy contribution, typically modeled as a log-normal distribution characterized by a median $e^{\mu}$ and a dispersion $\sigma_{\rm host}$. These quantities introduce degeneracies with $\Omega_b h^2$ and other parameters of the standard cosmological model. In order to alleviate these degeneracies, cosmological model-independent approaches are generally used, such as cosmography and non-parametric reconstructions of $H(z)$ (see, e.g., Refs. \cite{Sales:2025shu,ran2024cosmology,gomez2023neural,lemos2023cosmological}).

However, the currently available sample of localized FRBs is still not sufficiently precise to fully break the degeneracies between $\Omega_b h^2$, the host-galaxy contribution, and other cosmological quantities. In particular, uncertainties in the host-galaxy DM distribution propagate into the inference of the baryon density and limit the constraining power of FRB data alone when no additional information on the background expansion is provided. A useful strategy to mitigate this limitation is to reconstruct the cosmic expansion history in a non-parametric way through an artificial neural network (ANN). In this context, we adopt the Reconstruct Function with ANN (\texttt{ReFANN}) method developed by Wang et al. \cite{Wang:2019vxv}, which is trained using cosmic chronometer (CC) measurements to obtain a non-parametric reconstruction of $H(z)$. This reconstructed function is then incorporated into the theoretical mean IGM dispersion measure, while the cosmological parameter constraints are derived exclusively from localized FRB observations through a Markov chain Monte Carlo (MCMC) analysis. This procedure helps reduce degeneracies in the FRB likelihood and improves the precision of cosmological constraints that can be extracted from the current FRB sample. 

This work aims to probe the physical bayrion density from real FRB data, assuming a non-parametric reconstruction of the expansion history $H(z)$. Besides, we generated a mock catalog of 2000 synthetic FRBs to forecast future improvements, using a detection threshold that mimics the selection effects of the CHIME radio telescope. For the sake of comparison, we also used the Planck 2018 results \cite{aghanim2020planck} and Big Bang nucleosynthesis (BBN) \cite{cooke2018one} for comparing the values of $\Omega_b h^2$ constrained with our approach. In this way, FRBs provide a complementary low-redshift probe of baryons, enabling consistency tests with early-Universe measurements.

This paper is organized as follows. In Sec. \ref{sec2}, we present the main characteristics of the FRB theory. The non-parametric reconstruction method is discussed in Sec. \ref{rec_method}, where details of the neural network training procedure are introduced. In Sec. \ref{sec:methods}, we present the data and Bayesian inference procedure, specifying the priors and likelihood employed in our MCMC routine. The main results for the FRB dataset and the mock catalog are discussed in Sec. \ref{sec6}. Finally, in Sec. \ref{conclusion}, we present our final remarks and highlight the findings obtained in this work.


\section{Fast radio bursts}
\label{sec2}

The FRB population includes both apparently one-off events and repeating sources, suggesting diverse progenitors or emission channels. Some repeaters, such as FRB 121102, show extreme propagation effects and complex temporal and spectral behavior, indicating highly magnetized local environments. Observationally, FRBs exhibit strong dispersion, scattering, polarization, and in some cases large rotation measures, making them sensitive probes of the interstellar and intergalactic media. Their short durations and high brightness temperatures point to coherent emission processes. As new wide-field radio facilities continue to increase the detection rate, FRBs are becoming powerful tools for both astrophysical studies and cosmological applications. For general reviews, see Refs. \cite{petroff2019fast,cordes2019fast}.

A key observable of FRBs is their dispersion measure (DM), which plays a central role in extracting cosmological information from FRB observations. The observed DM by a radio telescope reflects the integrated contribution of free electrons along the line of sight, including the host galaxy (${\rm DM}_{\textrm{host}}$), the intergalactic medium (${\rm DM}_{\mathrm{IGM}}$), and the Milky Way (${\rm DM}_{\mathrm{MW}}$). It can be expressed as \cite{Petroff:2014taa}:
\begin{equation}
{\rm DM}_{\mathrm{obs}}(z)= {\rm DM}_{\mathrm{MW}}+{\rm DM}_{\mathrm{IGM}}+\frac{{\rm DM}_{\textrm{host}}}{1+z}\;,
\label{e1}
\end{equation}
where ${\rm DM}_{\rm MW} = {\rm DM}_{\rm MW,ISM}+{\rm DM}_{\rm MW,halo}$ with DM$_{\textrm{MW,ISM}}$ and DM$_{\textrm{MW,halo}}$ being the galactic interstellar medium and the galactic halo, respectively. The factor $(1+z)$ in \eqref{e1} comes from the cosmic dilation.

A time delay occurs in a detected pulse due to interactions with the components of Eq. \eqref{e1}, which also causes dispersion in the signal's frequencies. This difference is related to the free electrons density ($n_e$) present along the line of sight ($l$) to the FRB (see, e.g., \cite{Petroff:2019tty}) and expressed by the DM parameter. In the cosmological application of the FRBs, we should subtract any influence of the Milky Way and use exclusively the observed extragalactic DM. Consequently, from Eq. \eqref{e1}, we can write
\begin{align}
    {\rm DM}_{\rm ext}(z) &\equiv {\rm DM}_{\mathrm{obs}}(z)-{\rm DM}_{\mathrm{MW}}\nonumber\\
    &={\rm DM}_{\mathrm{IGM}}+\frac{{\rm DM}_{\textrm{host}}}{1+z}\;.
    \label{DM_ext_th}
\end{align}
%



The phenomenological probability distribution function (PDF) of $\rm DM_{IGM}$ follows an asymmetric distribution around the mean relation predicted by Macquart et al.~\cite{Macquart:2020lln}. It is defined in terms of the dimensionless variable $\Delta$ as
\begin{equation}
P_{\text {IGM}}(\Delta)=A \Delta^{-\beta} \exp \left[-\frac{\left(\Delta^{-\alpha}-C_0\right)^2}{2 \alpha^2 \sigma_{\text {IGM }}^2}\right], \quad \Delta>0
\label{pdf_distri}
\end{equation}
for which $A$ is a normalization constant and $C_0$ is determined by requiring $\left\langle\Delta\right\rangle=1$. For the current analysis, we adopt $\alpha=\beta=3$ and $\sigma_{\rm IGM}=Fz^{-0.5}$ with $F=0.32$, following \cite{Macquart:2020lln}. The variable $\Delta$ quantifies fluctuations in the free-electron density along different lines of sight and is defined through
\begin{equation}
\Delta \equiv \frac{\mathrm{DM}_{\mathrm{IGM}}}{\left\langle\mathrm{DM}_{\mathrm{IGM}}\right\rangle}=\frac{\mathrm{DM}_{\mathrm{ext}}-(1+z)^{-1}\mathrm{DM}_{\mathrm{host}}}{\left\langle\mathrm{DM}_{\mathrm{IGM}}\right\rangle}~.
\label{delta_igm}
\end{equation}

The average DM for the DM$_{\rm IGM}$ in terms of the redshift is described by the Macquart relation \cite{Macquart:2020lln}, whose form is
\begin{equation}
\langle {\rm DM}_{\rm IGM}(z) \rangle = A \Omega_b h^2 \int_0^z \frac{\left(1+z^{\prime}\right) x_e\left(z^{\prime}\right)}{H\left(z^{\prime}\right)} d z^{\prime}~,
\label{e2}
\end{equation}
with $A=3\times 10^{4}cf_{\rm IGM}/(8 \pi G m_p)$, where $G$ is Newton's gravitational constant, $f_{\rm IGM}=0.83$ is the baryon mass fraction present in the IGM \cite{shull2012baryon}, and $H(z)$ is the Hubble parameter to be reconstruted via neural network. The function $x_e(z)$ stands for the free electron fraction, and it is given by $x_{\rm e}(z)=Y_{\rm H} x_{\rm e, H}(z)+\frac{1}{2}Y_{\rm He} x_{\rm e, He}(z)$, with $Y_{\mathrm{H}}=3/4$ and $Y_{\mathrm{He}}=1/4$ being the mass fractions of hydrogen and helium. Here, $x_{\mathrm{e}, \mathrm{H}}(z)$ and $x_{\mathrm{e}, \mathrm{He}}(z)$ express the ionization fractions of hydrogen and helium, respectively. Since hydrogen and helium are fully ionized at $z<3$, we consider $x_{\rm e, H}(z)=x_{\rm e, He}(z)=1$ in our analysis \cite{Meiksin:2007rz,becker2011detection},  which provides $x_{\rm e}=7/8$.

The host-galaxy contribution, $\mathrm{DM}_{\mathrm{host}}$, originates from free electrons within the FRB’s host galaxy and depends on factors such as galaxy type, inclination, star formation activity, and the burst’s position within the host. These effects lead to a broad and asymmetric distribution. Hydrodynamical simulations, including IllustrisTNG, indicate that $\mathrm{DM}_{\mathrm{host}}$ is well described by a log-normal distribution across redshifts and galaxy populations \cite{zhang2020dispersion}. Accordingly, we model $\mathrm{DM}_{\mathrm{host}}$ using a log-normal form, as below: 
\begin{align} \label{p_host}
    P_{\rm host}(\mathrm{DM}_{\mathrm{host}}|z) =& \frac{1}{\sqrt{2\pi} \, \sigma_{\mathrm{host}} \, \mathrm{DM}_{\mathrm{host}}}\exp\left[-\frac{1}{2} \left(
\frac{\ln(\mathrm{DM}_{\mathrm{host}}) - \mu}{\sigma_{\mathrm{host}}}
\right)^2 \right]~,
\end{align}
where $\mu = \langle {\rm DM}_{\rm host} \rangle$ and $\sigma_{\rm host}$ are the distribution parameters. In this work, both parameters will be considered as free parameters.

\section{Non-parametric reconstruction with ANN} \label{rec_method}

The Hubble parameter $H(z)$ usually obtained from the $\Lambda$CDM model can also be reconstructed from cosmic chronometer (CC) observations through an artificial neural network (ANN). In this context, Wang et al. \cite{Wang:2019vxv} developed the Reconstruct Function with ANN (\texttt{ReFANN}\footnote{\url{https://github.com/Guo-Jian-Wang/refann.git}}) Python package. This method learns how to relate redshift and the Hubble parameter by a functional relationship, where each neuron effectively associates a given redshift with a corresponding value of $H(z)$. The network reconstruction is based on observational data training and provides a smooth function that best approximates the underlying behavior of the data set.

The connection between CC redshifts and their corresponding Hubble parameter values is established through a sequence of hidden layers, where the majority of neurons are located. Each hidden layer receives an input vector from the previous layer, performs a linear transformation, and subsequently applies a nonlinear activation function. In the present analysis, the exponential linear unit (ELU) activation function~\cite{clevert2015fast} is adopted,
\begin{equation}
    f(x)=
    \begin{cases}
        x, & x>0 \\
        \alpha\left(e^{x}-1\right), & x \leqslant 0
    \end{cases},
\end{equation}
where $\alpha$ is a positive hyperparameter that controls the saturation of the activation function for negative inputs. In this work, we fix $\alpha = 1$. The output of each neuron depends on the trainable parameters of the network, namely the weights and biases, which are determined during the training process. Further details of the network architecture and parameterization can be found in Ref. \cite{Wang:2019vxv}.

The loss function is responsible to quantify the the discrepancy between the observed value of the Hubble parameter and the predicted values from the ANN. Once this function is minimized, the ANN hyperparameters are also optimized and a new updated interaction occurs. After successive interactions, the network is trained as the minimal value of the loss function is reached and the ANN is trained with respect to the input dataset. The absolute error loss function is commonly chosen for that purpose \cite{Dialektopoulos:2023dhb},
\begin{equation}
    \mathrm{L}_1 =
    \sum_i
    \left|
        H_{\mathrm{obs}}(z_i) -
        H_{\mathrm{pred}}(z_i)
    \right|,
\end{equation}
being $H_{\mathrm{obs}}$ the observed value of the Hubble parameter and $H_{\mathrm{pred}}$ the ANN-predicted value obtained after each interaction. 

This approximate function is constructed by associating each redshift $z$ with the corresponding Hubble parameter and accounting for uncertainties in the $H(z)$ data. The implementation used in the original work, Ref. \cite{Wang:2019vxv}, is also adopted here, where we use  $3\times10^4$ as the total of interactions, being this number enough to reach the maximum reduction of the loss function. The learning rate starts at $0.01$ and is reduced as the interaction number increases.

\begin{figure}[h]
    \centering    \includegraphics[width=0.6\linewidth]{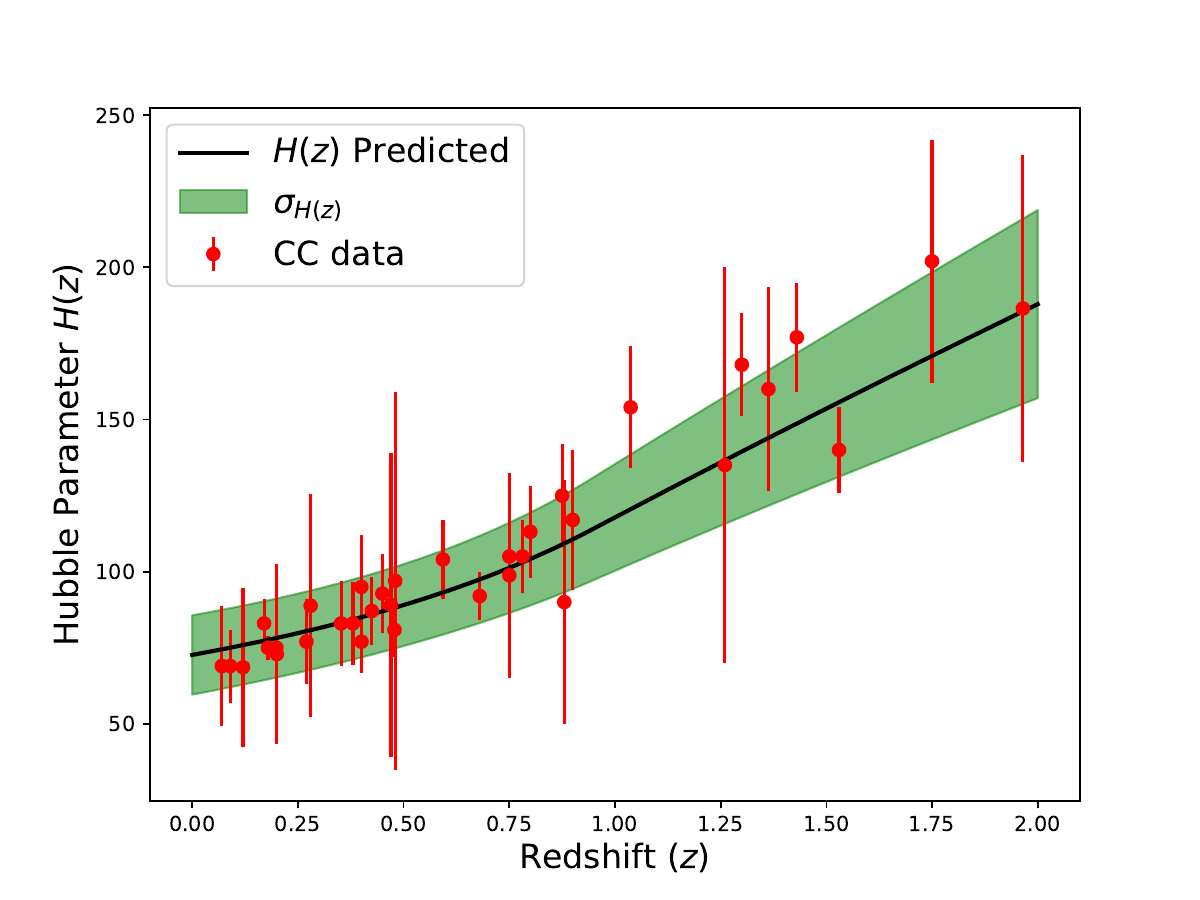}
    \caption{The reconstructed Hubble parameter as a function of redshift. The \texttt{ReFANN} result is depicted by the solid black curve, and the associated $1\sigma$ uncertainty given by the shaded green band. The red points and the error bars correspond to CC measurements.}
    \label{fig:h_rec}
\end{figure}

To determine an appropriate network architecture, we evaluate the statistical risk associated with different configurations of hidden layers. This quantity measures the expected prediction error and is defined as the combination of squared bias and variance terms \cite{Wasserman:2001ng},
\begin{align}
\text{risk} &= \sum_{i=1}^{N}\text{Bias}_i^2 + \sum_{i=1}^{N}\text{Variance}_i \nonumber\\
&= \sum_{i=1}^{N}\left[H(z_i)-\bar{H}(z_i)\right]^2 + \sum_{i=1}^{N}\sigma^2\!\left(H(z_i)\right)\,,
\end{align}
where $\bar{H}(z_i)$ represents the fiducial model and $N$ corresponds to the number of measurements in the $H(z_i)$ dataset. In the current work, we use the \texttt{ReFANN} code in its default configuration with the number of hidden layers set to 4096 neurons. Fig. \ref{fig:h_rec} depicts the reconstruction of the late-Universe expansion history from 35 CC\footnote{In the original work \cite{Wang:2019vxv}, 31 CC data are used.} measurements. Note that the reconstructed curve follows closely the CC observations, demonstrating that the neural network successfully captures the redshift evolution of the Hubble parameter within the range probed by the data.

\section{Data and Bayesian inference}
\label{sec:methods}

To reconstruct the late-Universe expansion history, we use 35 CC measurements in the redshift range $0.07 < z < 1.965$, which provide model-independent estimates of $H(z)$ based on the differential age evolution of massive galaxies \cite{Hu:2024big}. To perform the inference, the FRB dataset used in this analysis consists of 131 observations compiled from the catalogs presented in Refs. \cite{Sales:2025shu,Jia:2025kvp,Santos:2026qnl}. The $\mathrm{DM}_{\rm MW}$ is estimated using the NE2001 electron density model \cite{Cordes:2002wz}. Here, we utilize both repeating and non-repeating FRBs. It is worth stressing that the event FRB 20220319D exhibits a $\rm DM_{MW}$ larger than $\rm DM_{obs}$, which is physically inconsistent since the galactic contribution cannot exceed the total DM. Hence, such an event is excluded from the analysis, although it is still listed in Table \ref{tab1}. The complete FRB dataset is presented in Appendix \ref{append_a}, Table \ref{tab1}.

For the FRB sample, the likelihood is defined from the probability distribution of the extragalactic DM. This distribution is obtained by convolving the IGM contribution with a log-normal model describing the host-galaxy DM. In this way, the likelihood incorporates the main astrophysical uncertainties associated with FRB observations and forms the central ingredient of our analysis. Thus, we use the following likelihood \cite{Zhang:2025wif}: 
	\begin{align}
		P\left(\mathrm{DM}_{\text {ext}}\right)=&\frac{1}{\left\langle\mathrm{DM}_{\mathrm{IGM}}\right\rangle}\int_0^{\mathrm{DM}_{\text {ext }}(1+z)} P_{\text {host }}\left(\mathrm{DM}_{\text {host}}\right) \nonumber\\
		&\times P_{\rm IGM}\left(\frac{\mathrm{DM}_{\text {ext}}-(1+z)^{-1}\mathrm{DM}_{\text {host}}}{\left\langle\mathrm{DM}_{\mathrm{IGM}}\right\rangle}\right) d \mathrm{DM}_{\text {host}} .
		\label{prop5}
	\end{align}

Since all events are independent, the combined likelihood of the FRBs sample is simply the product of the individual likelihoods:
\begin{equation}
\mathcal{L}_{\rm total} = \prod_i P_i\left(\mathrm{DM}_{\text {ext}}\right)\;,
\end{equation}
and this product is performed for every FRB listed in Appendix \ref{append_a}, Table \ref{tab1}.

The parameter inference is carried out within a Bayesian framework using nested sampling techniques. Our analysis aims to assess the constraining power of FRBs by adopting a non-parametric reconstruction of the Hubble parameter. The parameters to be constrained in this analysis are
$\{\Omega_{\rm b}h^2, e^{\mu}, \sigma_{\rm host}\}$.
The corresponding prior distributions are summarized in Table \ref{tab_results}. We adopt broad, uninformative priors to minimize the impact of prior assumptions on the results.

The Hubble parameter $H(z)$ is first reconstructed from CC data using the \texttt{ReFANN} algorithm (see Section \ref{rec_method}). The reconstructed function is then employed in the FRB likelihood to constrain the parameters $\{\Omega_{\rm b}h^2, e^{\mu}, \sigma_{\rm host}\}$.

Posterior distributions and Bayesian evidence are computed using the nested sampling algorithm \texttt{MLFriends} implemented in the \texttt{UltraNest} package \cite{buchner2016statistical,buchner2019collaborative,buchner2021ultranest}. Marginalized constraints, confidence intervals, and contour plots are obtained with the \texttt{GetDist} package \cite{lewis2019getdist}.

\section{Results and discussions}\label{sec6}

In our Bayesian inference routine, the parameters $\{\Omega_{\rm b}h^2, e^{\mu}, \sigma_{\rm host}\}$ are constrained using real observations of FRBs. However, as a complementary analysis, a mock catalog of 2000 FRBs was generated to predict improvements in future FRB observations. This mock was built following recent studies of FRBs (see, e.g., \cite{chen2024formation,zhang2025revisiting}), employing an intrinsic formation rate, a luminosity function, and observed flux density, with a detection threshold that mimics CHIME selection effects. The DM are drawn using the Monte Carlo approach as explained in Appendix \ref{Ap.monte}.   

Fig. \ref{fig:1} shows the marginalized posterior distributions obtained
from the FRB samples for the parameters $\Omega_{\rm b}h^2$, $e^{\mu}$, and
$\sigma_{\rm host}$. The diagonal panels display the one-dimensional
marginalized posteriors, while the off-diagonal panels show the corresponding
two-dimensional credible regions at the 68\% and 95\% levels. For the real FRB
sample, the posteriors are relatively broad but well localized, indicating that the current data already provide meaningful constraints on the baryon density and on the statistical properties of the host-galaxy contribution.

\begin{figure}[h!]
    \centering    \includegraphics[width=0.7\linewidth]{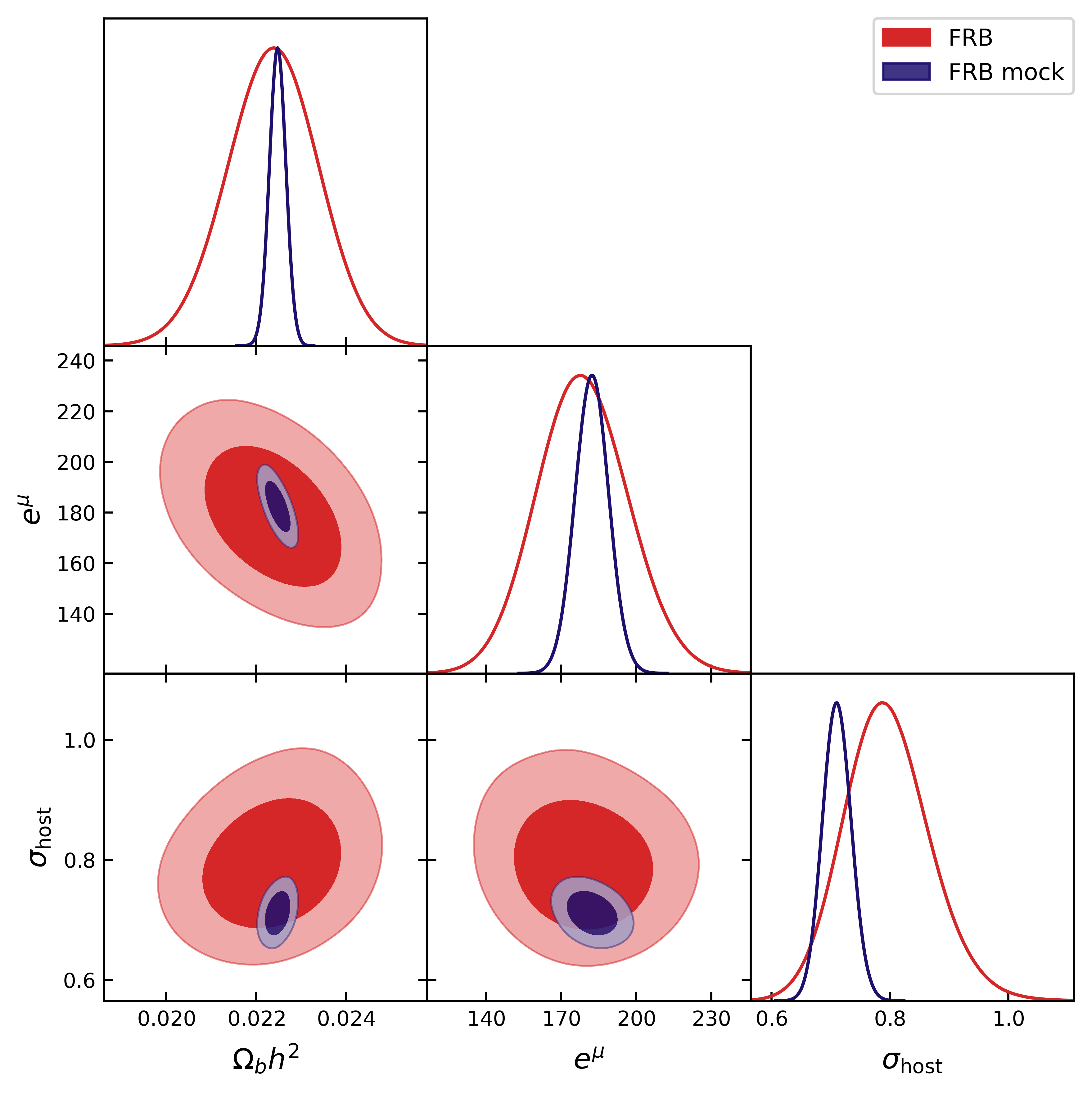}
    \caption{Corner plot showing the marginalized posterior distributions for $\Omega_{\rm b}h^2$, $e^{\mu}$, and $\sigma_{\rm host}$ inferred from the real FRB sample and the mock FRB catalog. The diagonal panels display the one-dimensional marginalized posterior distributions, while the off-diagonal panels show the corresponding two-dimensional credible regions at the $1\sigma$ and $2\sigma$ levels.}
    \label{fig:1}
\end{figure}

The mock FRB catalog leads to substantially narrower posterior distributions,
as expected from the increased statistical information encoded in the simulated
sample. Importantly, the mock constraints remain statistically consistent with
those inferred from the real data, with their posterior peaks lying within the
broader credible regions of the observed sample. This indicates that the mock
catalog preserves the main statistical trends of the current FRB population,
while illustrating the level of improvement that can be achieved with larger
and better-characterized FRB samples.


The two-dimensional marginalized contours also reveal the correlation structure
among the inferred parameters. The dominant degeneracy occurs between
$\Omega_{\rm b}h^2$ and $e^\mu$, for which we find a moderate anti-correlation
in the real FRB sample, with Pearson coefficient $r\simeq -0.43$. This behavior
has a direct physical interpretation: increasing $\Omega_{\rm b}h^2$ raises the
mean IGM contribution to the extragalactic DM, and this can be
partially compensated by lowering the median host-galaxy contribution $e^\mu$.
Thus, the anti-correlation reflects the fact that the observed FRB DM constrains the sum of cosmological and astrophysical contributions,
rather than each component independently.

For the mock FRB catalog, the same degeneracy becomes more pronounced, with
$r\simeq -0.66$ for the $\Omega_{\rm b}h^2$--$e^\mu$ pair. This stronger
anti-correlation is consistent with the narrower posterior contours obtained
for the mock sample: as the statistical uncertainties decrease, the remaining
allowed region becomes more concentrated along the intrinsic degeneracy
direction between the IGM and host-galaxy DM contributions. The remaining correlations are substantially weaker: for the real FRB sample we
obtain $r\simeq 0.18$ for $\Omega_{\rm b}h^2$--$\sigma_{\rm host}$ and
$r\simeq -0.06$ for $e^\mu$--$\sigma_{\rm host}$. In the mock case, these
increase mildly to $r\simeq 0.39$ and $r\simeq -0.33$, respectively, but remain
secondary compared with the dominant $\Omega_{\rm b}h^2$--$e^\mu$ degeneracy.


The priors adopted in the Bayesian inference and the resulting marginalized
constraints are summarized in Table \ref{tab_results}. For all parameters, we
adopt broad uniform priors, chosen to be weakly informative over the physically
relevant region of parameter space.

\begin{table}[h!]
\centering
\begin{tabular}{lcccc}
\hline\hline
Parameter & Prior type & Prior range & FRB & FRB mock \\
\hline
$\Omega_{\rm b} h^2$ 
& Uniform 
& $[0.005,\,0.1]$ 
& $0.02236^{+0.00090}_{-0.00090}$ 
& $0.02248^{+0.00018}_{-0.00018}$ \\

$e^{\mu}\,[\mathrm{pc}\,\mathrm{cm}^{-3}]$ 
& Uniform 
& $[20,\,300]$ 
& $178.15^{+16.51}_{-16.97}$ 
& $182.36^{+6.83}_{-6.48}$ \\

$\sigma_{\rm host}$ 
& Uniform 
& $[0.2,\,2]$ 
& $0.794^{+0.064}_{-0.067}$ 
& $0.711^{+0.024}_{-0.025}$ \\

\hline\hline
\end{tabular}
\caption{Uniform priors and marginalized posterior constraints obtained from
the real FRB sample and the mock FRB catalog. The quoted values correspond to
the marginalized central values and the associated $68\%$ credible intervals.}
\label{tab_results}
\end{table}

For the real FRB sample, we obtain
$e^{\mu}=178.15^{+16.51}_{-16.97}~\mathrm{pc}\,\mathrm{cm}^{-3}$ and
$\sigma_{\rm host}=0.794^{+0.064}_{-0.067}$. These constraints correspond to
relative uncertainties of approximately $9\%$ and $8\%$, respectively. The
mock FRB catalog yields significantly tighter constraints,
$e^{\mu}=182.36^{+6.83}_{-6.48}~\mathrm{pc}\,\mathrm{cm}^{-3}$ and
$\sigma_{\rm host}=0.711^{+0.024}_{-0.025}$, corresponding to relative
uncertainties of approximately $4\%$ and $3\%$. Therefore, the mock sample
reduces the allowed parameter volume while recovering host-galaxy parameters
that remain statistically compatible with those inferred from the real data.
This behavior indicates that the inference framework responds consistently to
an increase in statistical constraining power, as expected for larger and/or
higher-quality FRB samples.

These results improve upon previous FRB-based estimates of the host-galaxy
contribution, particularly because both $e^\mu$ and $\sigma_{\rm host}$ are
simultaneously inferred rather than fixed or indirectly estimated. For example,
\cite{yang2025constraining} obtained
$e^{\mu}=103^{+68}_{-48}~\mathrm{pc}\,\mathrm{cm}^{-3}$ using scattering
times as an estimator of $\mathrm{DM}_{\rm host}$, without explicitly
constraining the full host-distribution parameters. In turn,
\cite{james2022measurement} reported
$e^{\mu}=186^{+59}_{-48}~\mathrm{pc}\,\mathrm{cm}^{-3}$ assuming a
log-normal host distribution but fixing the scatter to
$\sigma_{\rm host}=3.5$. More recently, \cite{Sales:2025shu} found
$e^\mu=122.21^{+15.97}_{-17.10}~\mathrm{pc}\,\mathrm{cm}^{-3}$ and
$\sigma_{\rm host}=0.81^{+0.08}_{-0.09}$ using a cosmographic approach,
corresponding to relative uncertainties of approximately $12\%$ and $10\%$,
respectively. Compared with these analyses, our real FRB sample provides a more precise
determination of the median host contribution, with an uncertainty smaller by
a factor of roughly $2$--$4$ relative to earlier estimates. The constraint on
$\sigma_{\rm host}$ is also improved with respect to the cosmographic result
of \cite{Sales:2025shu}, while remaining broadly consistent with it. This
agreement is important because $\sigma_{\rm host}$ controls the intrinsic
width of the host-galaxy DM distribution and therefore directly affects the
separation between astrophysical and intergalactic contributions to the total DM.

\begin{figure}[h]
    \centering
    \includegraphics[width=0.49\linewidth]{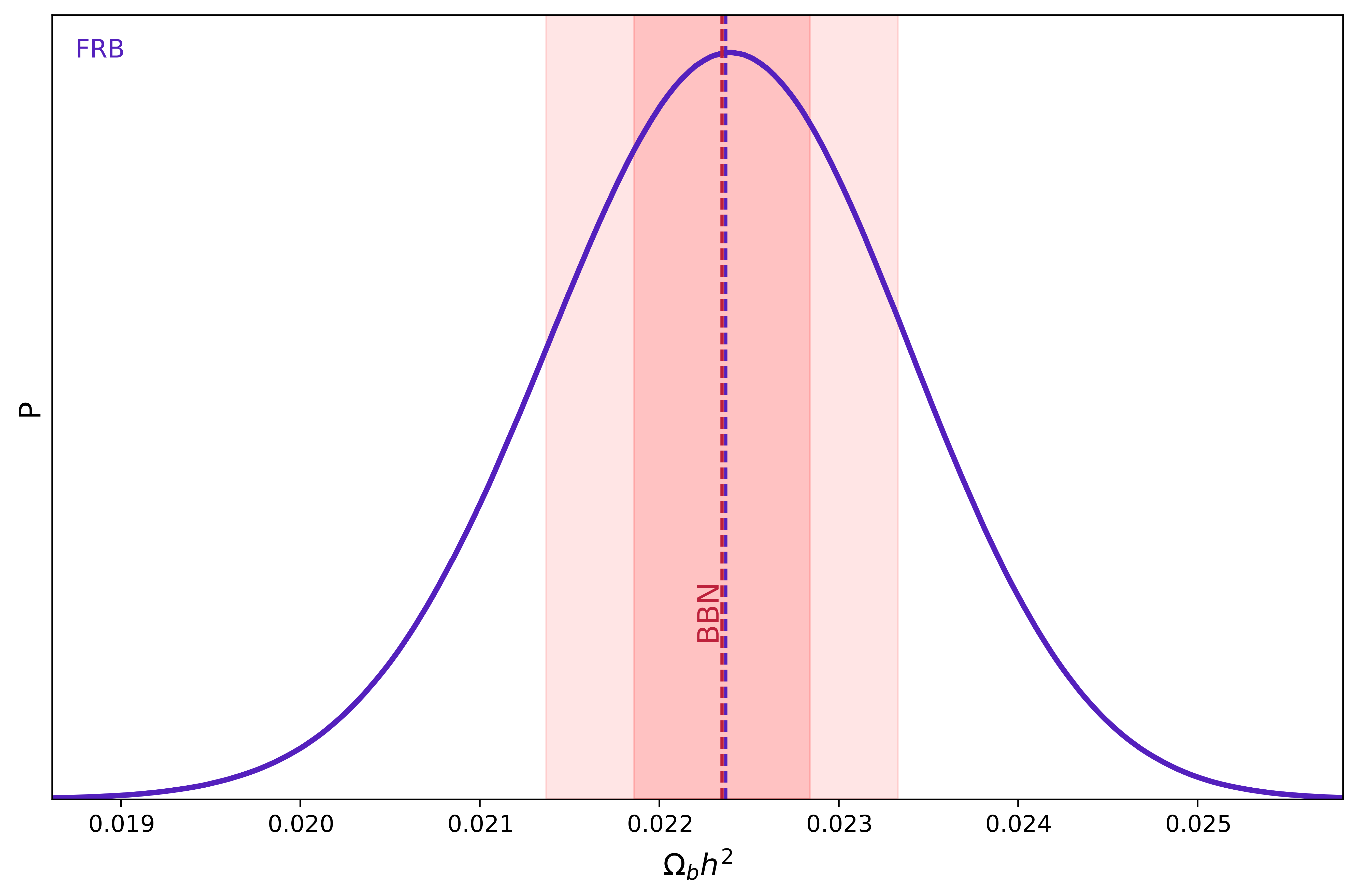}
    \includegraphics[width=0.49\linewidth]{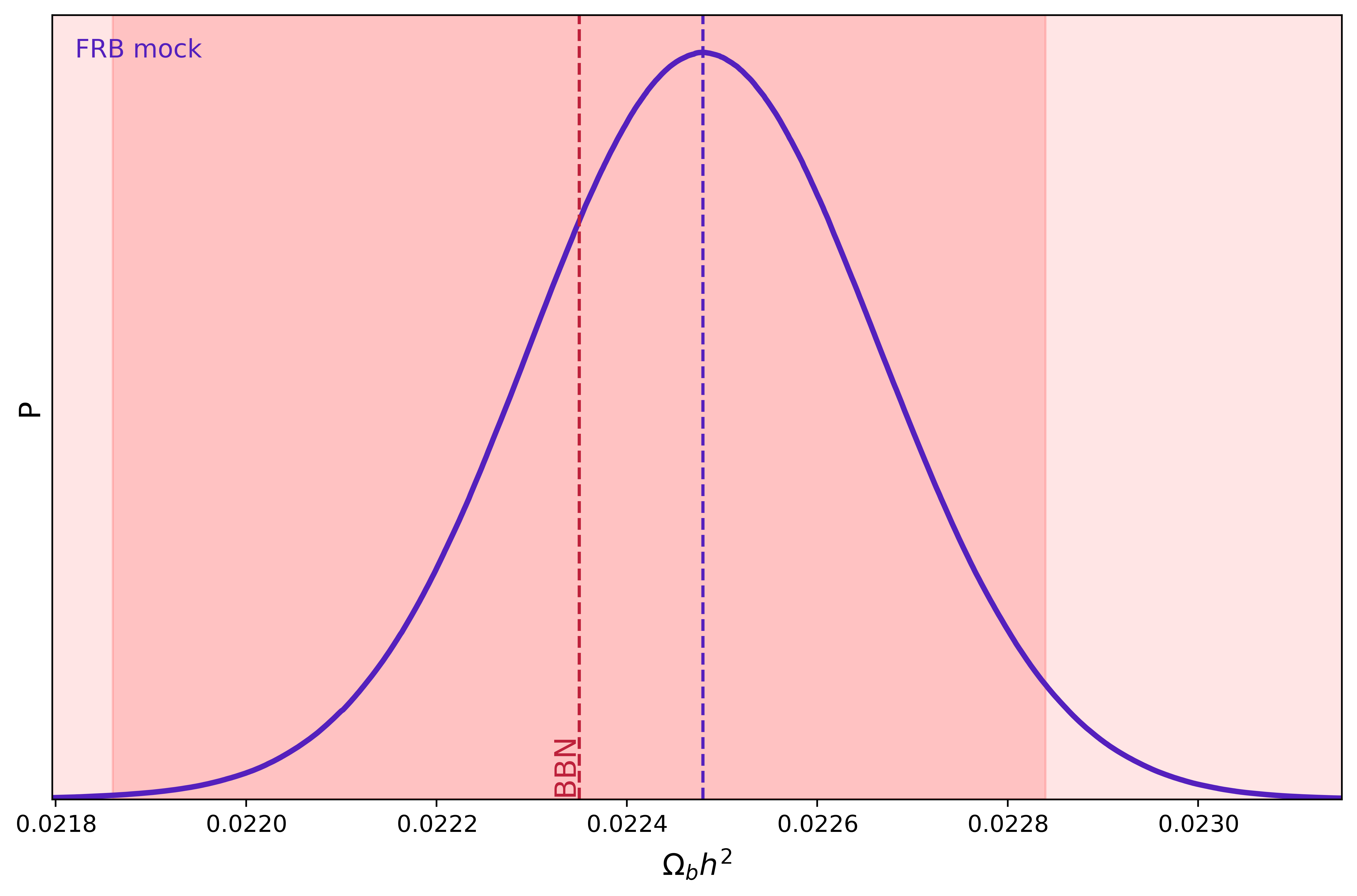}
    \caption{Posterior distributions of the baryon density parameter $\Omega_{\rm b}h^2$ inferred from the real FRB sample (left panel) and the mock FRB catalog (right panel). The purple dashed line indicates the FRB posterior central value, while the red dashed line marks the BBN mean $\Omega_{\rm b}h^2=0.02235$. The dark and light red shaded bands denote the corresponding $1\sigma$ and $2\sigma$ BBN
    intervals, respectively.}
    \label{fig:2}
\end{figure}

The posterior distributions of $\Omega_{\rm b}h^2$ inferred from the real and
mock FRB samples are compared with the BBN determination in
Fig. \ref{fig:2}. The BBN constraint,
$\Omega_{\rm b}h^2=0.02235\pm0.00049$ \cite{cooke2018one}, corresponding to
a relative uncertainty of $\sim 2.2\%$, is obtained from primordial
light-element abundances and therefore provides an independent early-Universe
reference for the baryon density. This comparison constitutes a
useful consistency test, since the FRB inference probes the baryonic content
through the DM accumulated along low-redshift lines of sight, whereas BBN constrains the same physical quantity from primordial nucleosynthesis.

For the real FRB sample, we obtain
$\Omega_{\rm b}h^2=0.02236\pm0.00090$, corresponding to a relative uncertainty
of $\sim 4\%$, in excellent agreement with the BBN value. The relative
difference between the posterior central value and the BBN mean is only
$\simeq 0.05\%$, corresponding to a negligible statistical offset. For the mock
FRB catalog, the inferred value
$\Omega_{\rm b}h^2=0.02248\pm0.00018$, corresponding to a relative uncertainty
of $\sim 0.8\%$, is also consistent with BBN, differing from the BBN mean by
only $\simeq 0.6\%$. In terms of the combined uncertainty, both estimates are
well within the $1\sigma$ region, indicating no evidence for a discrepancy
between the FRB-based inference and the standard BBN determination. The
comparison also illustrates the different constraining regimes of the real
and mock samples. The real FRB posterior is broader than the BBN reference band,
showing that current FRB data provide a compatible but less precise independent
constraint on $\Omega_{\rm b}h^2$. In contrast, the mock posterior is
substantially narrower, reflecting the increased statistical power expected
from larger and better-characterized FRB samples. The small shift of the mock
central value relative to BBN remains statistically insignificant and should be
interpreted within the assumptions used to generate the mock catalog.

\begin{figure}[h]
    \centering
    \includegraphics[width=0.49\linewidth]{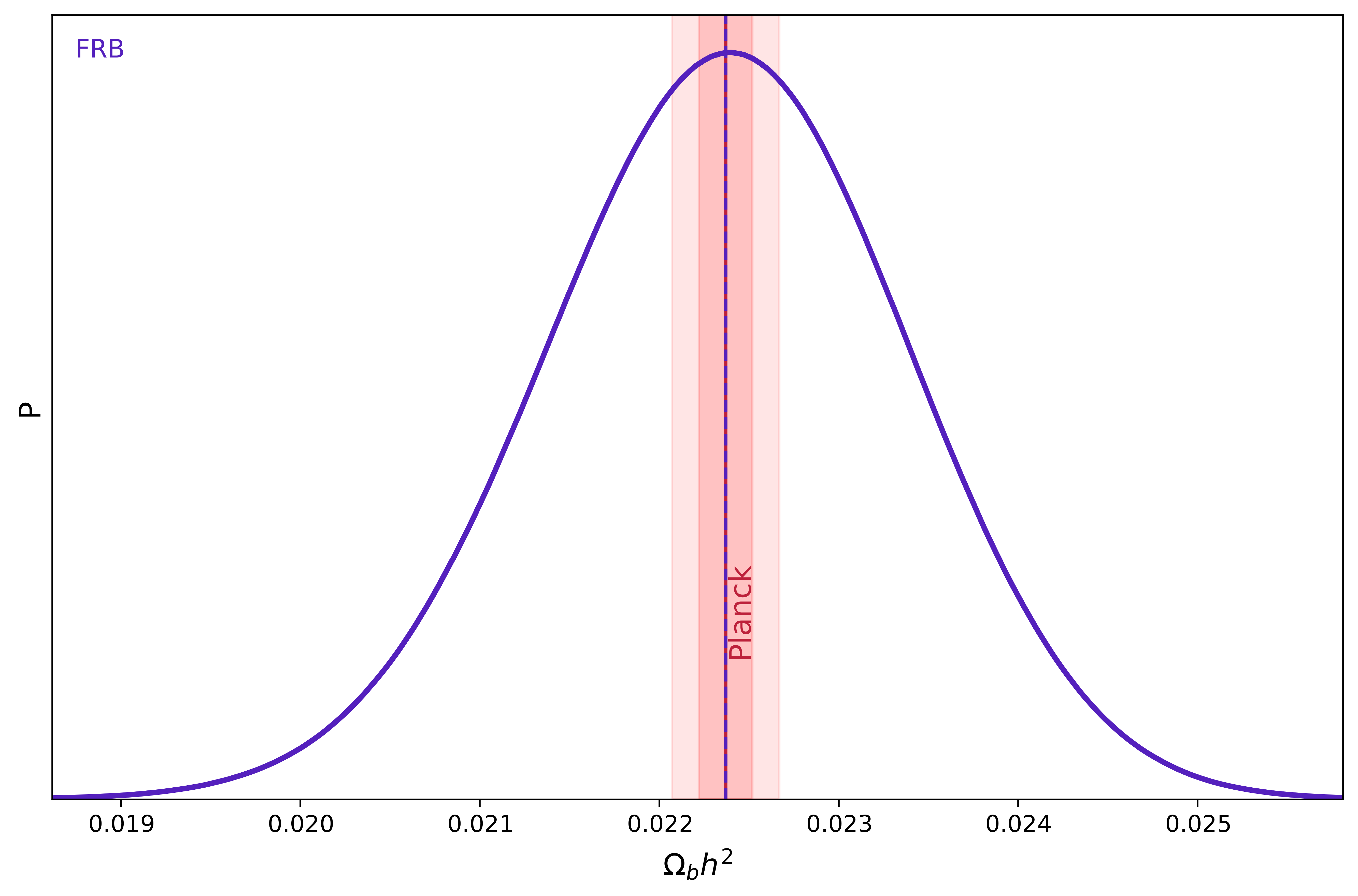}
    \includegraphics[width=0.49\linewidth]{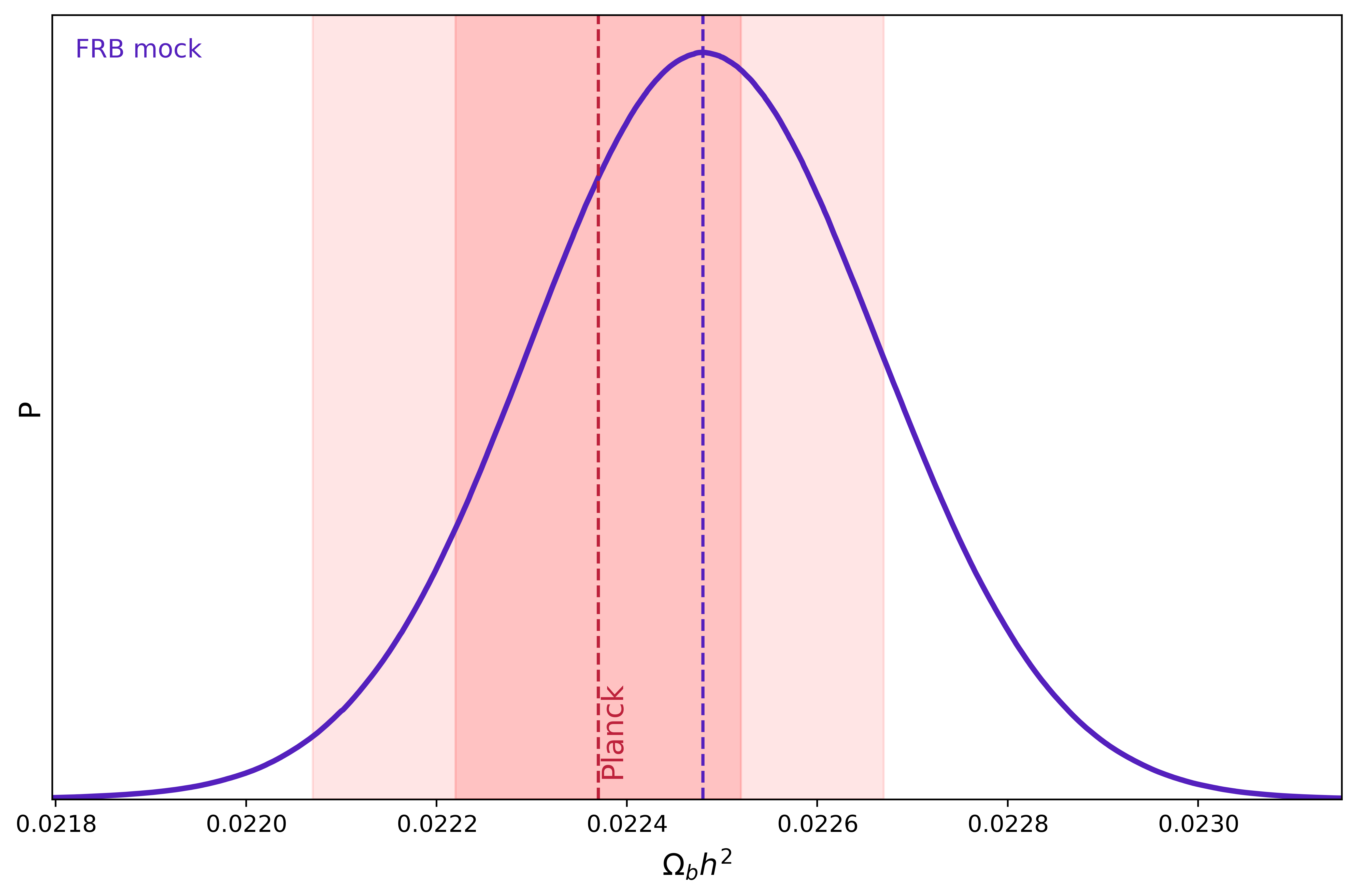}
    \caption{Same as Fig. \ref{fig:2}, but comparing the FRB posteriors with the Planck CMB constraint. The red dashed line and shaded bands indicate the Planck mean,
    $\Omega_{\rm b}h^2=0.02237$, and its $1\sigma$ and $2\sigma$ intervals,
    respectively.}
    \label{fig:2a}
\end{figure}

Fig. \ref{fig:2a} compares the posterior distributions of
$\Omega_{\rm b}h^2$ inferred from the real and mock FRB samples with the
Planck CMB constraint,
$\Omega_{\rm b}h^2=0.02237\pm0.00015$ \cite{aghanim2020planck}, corresponding to a relative uncertainty
of $\sim 0.7\%$, obtained within the standard
$\Lambda$CDM framework. This comparison provides a complementary consistency
test with respect to the CMB determination of the baryon density. In
the case of the real FRB sample, we obtain
$\Omega_{\rm b}h^2=0.02236\pm0.00090$, which is essentially coincident with
the Planck central value. The relative difference is only $\simeq 0.05\%$,
corresponding to a negligible offset when the uncertainties are combined.
This agreement indicates that the FRB-based inference of the baryonic content
at low redshift is fully consistent with the CMB predictions of the baryon density. For the mock FRB catalog, the posterior is considerably narrower and peaks at
$\Omega_{\rm b}h^2=0.02248\pm0.00018$, lying slightly above the Planck mean by
$\simeq 0.49\%$. Despite this small displacement, the result remains
statistically consistent with Planck, with a difference below the $1\sigma$
level when the uncertainties of both measurements are taken into account. Therefore, the mock result illustrates that future FRB samples may reach a precision
comparable to early-Universe constraints.

Overall, the comparisons with BBN and Planck reinforce the consistency of the
inferred FRB baryon density with standard cosmological expectations. The real
FRB posterior is broader than both the BBN and Planck reference intervals,
reflecting the current statistical and astrophysical limitations of FRB-only
constraints. Nevertheless, the use of a non-parametric neural network reconstruction of $H(z)$, implemented through the \texttt{ReFANN} framework, reduces the dependence on a specific background cosmological model and helps alleviate model-driven degeneracies in the FRB inference. This contributes to a more precise determination of $\Omega_{\rm b}h^2$ from FRB data.

In contrast, the mock posterior shows the potential of larger and better-characterized FRB samples to substantially reduce the allowed parameter volume, reaching a precision comparable to that of early-Universe probes. In this sense, the mock should be interpreted primarily as a forecasting and internal consistency test, rather than as a definitive prediction of the constraining power of future surveys.


\section{Conclusions}
\label{conclusion}

In this work, we investigated the potential of fast radio bursts as probes of
the cosmic baryon content by combining FRB data with a non-parametric reconstruction of the Universe expansion history. The reconstruction
of $H(z)$ was performed with the \texttt{ReFANN} code using observational
cosmic chronometer data, independently of the FRB sample. This reconstructed
expansion history was then used in the FRB likelihood to compute the contribution to the average DM$_{\rm IGM}$, thereby reducing the dependence of the analysis on a specific background cosmological model. In this sense, the role of \texttt{ReFANN} is to provide a flexible late-time reconstruction of
the expansion rate entering the FRB cosmological inference.

Within the Bayesian inference routine, we obtained competitive constraints on the baryon density parameter. For the real FRB sample, we found
$\Omega_{\rm b}h^2=0.02236\pm0.00090$, corresponding to a relative uncertainty
of approximately $4\%$. This value is in excellent agreement with both the BBN
determination, $\Omega_{\rm b}h^2=0.02235\pm0.00049$, and the Planck CMB
constraint, $\Omega_{\rm b}h^2=0.02237\pm0.00015$, with relative differences
of only $\simeq 0.05\%$ in both cases. These results indicate that the
FRB-based inference of the baryon density at low redshift is fully consistent
with independent early-Universe determinations.

We also explored the constraining power expected from a larger FRB sample by
constructing a mock catalog with 2000 events. For this mock sample, we obtained
$\Omega_{\rm b}h^2=0.02248\pm0.00018$, corresponding to a relative uncertainty
of approximately $0.8\%$. Although the mock posterior is slightly shifted with
respect to the BBN and Planck central values, by $\simeq 0.6\%$ and
$\simeq 0.49\%$, respectively, it remains statistically consistent with both
reference measurements. The substantial reduction in the posterior width
illustrates the potential of future large and well-characterized FRB samples
to approach the precision of early-Universe probes while providing an
independent late-time measurement of the baryon density.

In addition to baryon density, we simultaneously constrained the
parameters describing the host-galaxy contribution to the observed DM. For the median host contribution, we found
$e^\mu=178.15^{+16.51}_{-16.97}~\mathrm{pc}\,\mathrm{cm}^{-3}$ for the real
FRB sample, corresponding to a precision of approximately $9\%$. For the mock
catalog, this improves to
$e^\mu=182.36^{+6.83}_{-6.48}~\mathrm{pc}\,\mathrm{cm}^{-3}$, with a precision
of approximately $4\%$. These constraints are tighter than several previous ones
FRB-based estimates, particularly because the median host contribution is
inferred jointly with the baryon density and the host-distribution scatter,
rather than being fixed or indirectly estimated. The logarithmic scatter of the host-galaxy contribution was also constrained.
For the real FRB sample, we obtained
$\sigma_{\rm host}=0.794^{+0.064}_{-0.067}$, corresponding to a relative
uncertainty of approximately $8\%$. In the mock case, the constraint improves
to
$\sigma_{\rm host}=0.711^{+0.024}_{-0.025}$, corresponding to a precision of
approximately $3\%$. This improvement is particularly relevant because
$\sigma_{\rm host}$ encodes the intrinsic diversity of host environments and
therefore affects the separation between astrophysical and intergalactic
contributions to the observed DM. Accurately
constraining this parameter is essential for preventing biases in the inferred
cosmic baryon density.

Overall, our findings show that combining FRB DMs with a non-parametric reconstruction of the expansion history provides a robust pathway to constrain both cosmological and astrophysical parameters. The agreement with BBN and Planck demonstrates that the inferred FRB baryon density is consistent
with standard cosmological expectations. At the same time, the mock analysis shows that future FRB surveys can substantially reduce the allowed parameter volume. The method should therefore be regarded as a complementary late-time probe of the cosmic baryon density, whose ultimate precision will depend on the size and quality of localized FRB samples, the characterization of host-galaxy environments, the modeling of IGM scatter, and the control of selection effects.

{\acknowledgments}

A.R.Q., L.L.S., K.E.L.F., B.W.R. and R.A.B. thank the Para\'iba State Research Foundation (FAPESQ) for financial support.  A.R.Q. also acknowledges the support by CNPq under process number 310533/2022-8. K.E.L.F. and R.H.S. also thank CNPq for financial support.

\appendix
\label{append_a}
\section{Fast radio bursts dataset}

\begin{longtable}{lcccc}
\caption{\label{tab1} Fast radio bursts dataset.}\\
FRB (name) & Redshift $z$ & DM$_{\rm obs}$ (pc cm$^{-3}$) & DM$_{\rm MW}$ (pc cm$^{-3}$) & Reference \\
\endfirsthead

\caption[]{Fast radio bursts dataset (continued).}\\
FRB (name) & Redshift $z$ & DM$_{\rm obs}$ (pc cm$^{-3}$) & DM$_{\rm MW}$ (pc cm$^{-3}$) & Reference \\
\endhead

\multicolumn{5}{r}{Continued on next page} \\
\endfoot

\endlastfoot
20200120E & 0.0008 & 87.77 & 30 & \cite{Bhardwaj:2021xaa}\\
20181030A & 0.0039 & 103.5 & 40 & \cite{Bhardwaj:2021hgc,CHIMEFRB:2021srp}\\
20250316A & 0.0067 & 161.82 & 70 & \cite{CHIME:2025mlf} \\
20171020A & 0.00867 & 114.1 & 38 & \cite{Mahony:2018ddp}\\
20231229A & 0.0190 & 198.5 & 58.12 & \cite{CHIMEFRB:2025ggb}\\
20240210A & 0.0238 & 283.75 & 31 & \cite{Shannon:2024pbu}\\
20181220A & 0.027 & 209.4 & 125.8 & \cite{Bhardwaj:2023vha,CHIMEFRB:2021srp}\\
20231230A & 0.0298 & 131.4 & 61.51 & \cite{CHIMEFRB:2025ggb}\\
20181223C & 0.03024 & 112.51 & 19.91 & \cite{Bhardwaj:2023vha}\\
20190425A & 0.03122 & 128.16 & 48.75 & \cite{Bhardwaj:2023vha,CHIMEFRB:2021srp}\\
20180916B & 0.0337 & 348.76 & 200 & \cite{Gordon:2023cgw,CHIMEFRB:2021srp}\\
20230718A & 0.035 & 477 & 396 & \cite{glowacki2024h}\\
20240201A & 0.0427 & 374.5 & 38 & \cite{Shannon:2024pbu}\\
20220207C & 0.0430 & 262.38 & 79.3 & \cite{Law:2023ibd}\\
20211127I & 0.0469 & 234.83 & 42.5 & \cite{Gordon:2023cgw}\\
20201123A & 0.0507 & 433.55 & 251.93 & \cite{Rajwade:2022zkj}\\
20230926A & 0.0553 & 222.8 & 52.69 & \cite{CHIMEFRB:2025ggb}\\
20200223B & 0.06024 & 202.268 & 46 & \cite{Ibik:2023ugl,CHIMEFRB:2021srp}\\
20190303A & 0.064 & 222.4 & 26 & \cite{Michilli:2022bbs,CHIMEFRB:2021srp}\\
20231204A & 0.0644 & 221.0 & 29.73 & \cite{CHIMEFRB:2025ggb}\\
20231206A & 0.0659 & 457.7 & 59.13 & \cite{CHIMEFRB:2025ggb}\\
20210405I & 0.066 & 565.17 & 516.1 & \cite{Driessen:2023lxj}\\
20180814 & 0.068 & 189.4 & 87 & \cite{Michilli:2022bbs,CHIMEFRB:2021srp}\\
20231120A & 0.07 & 438.9 & 43.8 & \cite{DeepSynopticArrayTeam:2023iev,Sharma:2024fsq,Connor:2024mjg}\\
20231005A & 0.0713 & 189.4 & 33.37 & \cite{CHIMEFRB:2025ggb}\\
20190418A & 0.07132 & 184.5 & 70.1 & \cite{Bhardwaj:2023vha,CHIMEFRB:2021srp}\\
20211212A & 0.0715 & 206.0 & 27.1 & \cite{Gordon:2023cgw}\\
20231123A & 0.0729 & 302.1 & 89.76 & \cite{CHIMEFRB:2025ggb}\\
20220912A & 0.0771 & 219.46 & 115 & \cite{DeepSynopticArrayTeam:2022rbq,Zhang:2023eui}\\
20231011A & 0.0783 & 186.3 & 70.36 & \cite{CHIMEFRB:2025ggb}\\
20220509G & 0.0894 & 269.53 & 55.2 & \cite{Law:2023ibd}\\
20230124 & 0.0940 & 590.6 & 38.5 & \cite{DeepSynopticArrayTeam:2023iev,Sharma:2024fsq,Connor:2024mjg}\\
20201124A & 0.098 & 413 & 123 & \cite{Gordon:2023cgw,Lanman:2021yba}\\
20230708A & 0.105 & 411.51 & 50 & \cite{Shannon:2024pbu}\\
20231223C & 0.1059 & 165.8 & 47.9 & \cite{CHIMEFRB:2025ggb}\\
20191106C & 0.10775 & 333.4 & 25 & \cite{Ibik:2023ugl,CHIMEFRB:2021srp}\\
20231128A & 0.1079 & 331.6 & 25.05 & \cite{CHIMEFRB:2025ggb}\\
20230222B & 0.11 & 187.8 & 27.7 & \cite{CHIMEFRB:2025ggb}\\
20231201A & 0.1119 & 169.4 & 70.03 & \cite{CHIMEFRB:2025ggb}\\
20220319D & 0.0112 & 110.98 & 133.3 & \cite{Law:2023ibd} \\
20220914A & 0.1139 & 631.28 & 55.2 & \cite{Law:2023ibd}\\
20190608B & 0.1178 & 339 & 37 & \cite{Gordon:2023cgw,Hiramatsu:2022tyn}\\
20230703A & 0.1184 & 291.3 & 26.97 & \cite{CHIMEFRB:2025ggb}\\
20240213A & 0.1185 & 357.4 & 40.1 & \cite{Connor:2024mjg}\\
20240318A & 0.12 & 256.4 & 37 & \cite{Shannon:2024pbu}\\
202030222A & 0.1223 & 706.1 & 134.13 & \cite{CHIMEFRB:2025ggb}\\
20190110C & 0.1224 & 221.961 & 35.66 & \cite{Ibik:2023ugl,CHIMEFRB:2021srp}\\
20230628A & 0.1265 & 345.15 & 39.1 & \cite{DeepSynopticArrayTeam:2023iev,Sharma:2024fsq,Connor:2024mjg}\\
20240310A & 0.127 & 601.8 & 36 & \cite{Shannon:2024pbu}\\
20210807D & 0.1293 & 251.9 & 121.2 & \cite{Gordon:2023cgw}\\
20240114A & 0.13 & 527.65 & 49.7 & \cite{Kumar:2024svu}\\
20240209A & 0.1384 & 176.49 & 55.5 & \cite{Shah:2024ywp}\\
20210410D & 0.1415 & 578.78 & 56.2 & \cite{Gordon:2023cgw,Caleb:2023atr}\\
20230203A & 0.1464 & 420.1 & 36.29 & \cite{CHIMEFRB:2025ggb}\\
20231226A & 0.1569 & 329.9 & 145 & \cite{Shannon:2024pbu}\\
20230526A & 0.157 & 316.4 & 50 & \cite{Shannon:2024pbu}\\
20220920A & 0.158 & 314.99 & 40.3 & \cite{Law:2023ibd}\\
20200430A & 0.1608 & 380.25 & 27 & \cite{Gordon:2023cgw,Hiramatsu:2022tyn}\\
20210603A & 0.177 & 500.15 & 40 & \cite{Cassanelli:2023hvg}\\
20220529A & 0.1839 & 246.0 & 40.0 & \cite{gao2024measuring}\\
20230311A & 0.1918 & 364.3 & 92.39 & \cite{CHIMEFRB:2025ggb}\\
20220725A & 0.1926 & 290.4 & 31 & \cite{Shannon:2024pbu}\\
20121102A & 0.19273 & 557.0 & 188.4 & \cite{Gordon:2023cgw}\\
20221106A & 0.2044 & 343.8 & 35 & \cite{Shannon:2024pbu}\\
20240215A & 0.21 & 549.5 & 48.0 & \cite{Connor:2024mjg}\\
20230730A & 0.2115 & 312.5 & 85.18 & \cite{CHIMEFRB:2025ggb}\\
20210117A & 0.214 & 729.0 & 34.0 & \cite{Bhandari:2022ton}\\
20221027A & 0.229 & 452.5 & 47.2 & \cite{Connor:2024mjg}\\
20191001A & 0.234 & 506.92 & 44.7 & \cite{Gordon:2023cgw,Bhandari:2020cde}\\
20190714A & 0.2365 & 504.13 & 38 & \cite{Gordon:2023cgw,Hiramatsu:2022tyn,HESS:2021smp,Guidorzi:2020ggq}\\
20221101B & 0.2395 & 490.7 & 131.2 & \cite{DeepSynopticArrayTeam:2023iev,Sharma:2024fsq,Connor:2024mjg}\\
20220825A & 0.2414 & 651.24 & 79.7 & \cite{Law:2023ibd}\\
20190520B & 0.2418 & 1204.7 & 60.2 & \cite{Gordon:2023cgw}\\
20191228A & 0.2432 & 297.5 & 33 & \cite{Bhandari:2021pvj}\\
20231017A & 0.2450 & 344.2 & 64.55 & \cite{CHIMEFRB:2025ggb}\\
20221113A & 0.2505 & 411.4 & 91.7 & \cite{DeepSynopticArrayTeam:2023iev,Sharma:2024fsq,Connor:2024mjg}\\
20220307B & 0.2507 & 499.15 & 128.2 & \cite{Law:2023ibd}\\
20220831A & 0.262 & 1146.25 & 126.7 & \cite{Connor:2024mjg}\\
20231123B & 0.2625 & 396.7 & 40.2 & \cite{DeepSynopticArrayTeam:2023iev,Sharma:2024fsq,Connor:2024mjg}\\
20230307A & 0.2710 & 608.9 & 37.6 & \cite{DeepSynopticArrayTeam:2023iev,Sharma:2024fsq,Connor:2024mjg}\\
20221116A & 0.2764 & 640.6 & 132.3 & \cite{Sharma:2024fsq,Connor:2024mjg}\\
20220105A & 0.2785 & 583 & 22 & \cite{Gordon:2023cgw}\\
20210320C & 0.2796 & 384.8 & 42.2 & \cite{Gordon:2023cgw}\\
20221012A & 0.2846 & 441.08 & 54.4 & \cite{Law:2023ibd}\\
20240229A & 0.287 & 491.15 & 37.9 & \cite{Connor:2024mjg}\\
20190102C & 0.2913 & 363.6 & 57.3 & \cite{Gordon:2023cgw}\\
20220506D & 0.3004 & 396.97 & 89.1 & \cite{Law:2023ibd}\\
20230501A & 0.3010 & 532.5 & 125.6 & \cite{DeepSynopticArrayTeam:2023iev,Connor:2024mjg}\\
20230503E & 0.32 & 483.74 & 88 & \cite{Pastor-Marazuela:2025loc}\\
20180924B & 0.3214 & 361.42 & 40.5 & \cite{Gordon:2023cgw}\\
20231025B & 0.3238 & 368.7 & 48.67 & \cite{CHIMEFRB:2025ggb}\\
20230125D & 0.3265 & 640.08 & 88 & \cite{Pastor-Marazuela:2025loc}\\
20230626A & 0.3270 & 451.2 & 39.2 & \cite{DeepSynopticArrayTeam:2023iev,Sharma:2024fsq,Connor:2024mjg}\\
20180301A & 0.3304 & 536 & 152 & \cite{Gordon:2023cgw,Price:2019fmc}\\
20231220A & 0.3355 & 491.2 & 49.9 & \cite{Connor:2024mjg}\\
20211203C & 0.3439 & 635.0 & 63.4 & \cite{Gordon:2023cgw}\\
20220208A & 0.3510 & 437.0 & 101.6 & \cite{Sharma:2024fsq,Connor:2024mjg}\\
20220726A & 0.3610 & 686.55 & 89.5 & \cite{DeepSynopticArrayTeam:2023iev,Sharma:2024fsq,Connor:2024mjg}\\
20220717A & 0.36295 & 637.34 & 118.3 & \cite{Rajwade:2024ozu}\\
20230902A & 0.3619 & 440.1 & 34 & \cite{Shannon:2024pbu}\\
20200906A & 0.3688 & 577.8 & 36 & \cite{Gordon:2023cgw,Hiramatsu:2022tyn}\\
20240119A & 0.37 & 483.1 & 37.9 & \cite{Connor:2024mjg}\\
20220330D & 0.3714 & 468.1 & 38.6 & \cite{Sharma:2024fsq,Connor:2024mjg}\\
20190611B & 0.3778 & 321.4 & 57.8 & \cite{Gordon:2023cgw}\\
20220501C & 0.381 & 449.5 & 31 & \cite{Shannon:2024pbu}\\
20240208A & 0.39 & 260.2 & 98 & \cite{Shannon:2024pbu}\\
20230613A & 0.3923 & 483.51 & 30 & \cite{Pastor-Marazuela:2025loc}\\
20220204A & 0.4 & 612.2 & 50.7 & \cite{DeepSynopticArrayTeam:2023iev,Sharma:2024fsq,Connor:2024mjg} \\
20230712A & 0.4525 & 586.96 & 39.2 & \cite{DeepSynopticArrayTeam:2023iev,Sharma:2024fsq,Connor:2024mjg}\\
20230907D & 0.4638 & 1030.79 & 29 & \cite{Pastor-Marazuela:2025loc}\\
20181112A & 0.4755 & 589.27 & 42 & \cite{Gordon:2023cgw}\\
20231020B & 0.4775 & 952.2 & 34 & \cite{Pastor-Marazuela:2025loc}\\
20220310F & 0.4779 & 462.24 & 45.4 & \cite{Law:2023ibd} \\
20220918A & 0.491 & 656.8 & 41 & \cite{Shannon:2024pbu}\\
20231210F & 0.5 & 720.6 & 32 & \cite{Pastor-Marazuela:2025loc}\\
20190711A & 0.5220 & 593.1 & 56.4 & \cite{Gordon:2023cgw,Macquart:2020lln}\\
20230216A & 0.5310 & 828.0 & 38.5 & \cite{Sharma:2024fsq,Connor:2024mjg}\\
20230814B & 0.553 & 696.4 & 104 & \cite{Connor:2024mjg}\\
20230814A & 0.5535 & 696.4 & 104.9 & \cite{Connor:2024mjg} \\
20221219A & 0.5540 & 706.7 & 44.4 & \cite{DeepSynopticArrayTeam:2023iev,Sharma:2024fsq,Connor:2024mjg}\\
20190614D & 0.60 & 959.2 & 83.5 & \cite{Law:2020cnm,Hiramatsu:2022tyn}\\
20231010A & 0.61 & 442.59 & 41 & \cite{Pastor-Marazuela:2025loc}\\
20220418A & 0.6220 & 623.25 & 37.6 & \cite{Law:2023ibd}\\
20220224C & 0.6271 & 1140.2 & 52 & \cite{Pastor-Marazuela:2025loc}\\
20190523A & 0.6600 & 760.8 & 37 & \cite{Ravi:2019alc}\\
20220222C & 0.853 & 1071.2 & 56 & \cite{Pastor-Marazuela:2025loc}\\
20240123A & 0.968 & 1462.0 & 90.3 & \cite{Connor:2024mjg}\\
20221029A & 0.9750 & 1391.05 & 43.9 & \cite{DeepSynopticArrayTeam:2023iev,Sharma:2024fsq,Connor:2024mjg} \\
20220610A & 1.016 & 1458.1 & 30.9 & \cite{Ryder:2022qpg}\\
20230521B & 1.354 & 1342.9 & 138.8 & \cite{Shannon:2024pbu,Connor:2024mjg}\\
20240304B & 2.148 & 2458.2 & 28.1 & \cite{Caleb:2025uzd}\\
\end{longtable}

\section{Monte Carlo Simulation of Mock FRB Catalog}
\label{Ap.monte}

In this work, we generate a synthetic catalog of FRBs using a Monte Carlo
framework designed to reproduce the statistical and physical properties
inferred from observational analyses. The methodology closely follows the
population-modeling strategy adopted in recent studies of FRB formation-rate
evolution and luminosity functions (see, e.g.,
\cite{chen2024formation,zhang2025revisiting}).

We first compute the comoving distance,
\begin{equation}
D_c(z) =
\frac{c}{H_0}
\int_0^z \frac{dz'}{E(z')},
\end{equation}
where $c$ is the speed of light. We assume a spatially flat $\Lambda$CDM
fiducial cosmology with
$H_0 = 67.4 \, \mathrm{km\,s^{-1}\,Mpc^{-1}}$ and
$\Omega_{\rm m} = 0.315$ \cite{aghanim2020planck}, for which
\begin{equation}
E(z)=\sqrt{\Omega_{\rm m}(1+z)^3+1-\Omega_{\rm m}}.
\end{equation}
The luminosity distance is then given by
$D_L(z)=(1+z)D_c(z)$.

To determine the redshift distribution of the FRB cosmological population, we
use the differential comoving volume element
\begin{equation}
\frac{dV}{dz}
=
4\pi
\frac{c}{H_0}
\frac{D_c^2(z)}{E(z)}.
\end{equation}
Based on the findings of Ref.~\cite{chen2024formation}, the intrinsic FRB
formation rate is assumed to evolve as
\begin{equation}
\rho(z) \propto (1+z)^{-4.9}.
\end{equation}
Due to cosmological time dilation, the observable redshift probability density
function is taken as
\begin{equation}
p(z) \propto
\frac{\rho(z)}{1+z}
\frac{dV}{dz}.
\end{equation}
This distribution incorporates the astrophysical evolution $\rho(z)$, the
cosmological volume factor $dV/dz$, and the relativistic time-dilation factor
$(1+z)^{-1}$. Redshifts are sampled through rejection sampling in the interval
$0.001<z<2.5$.

To enhance the realism of the mock catalog, we adopt a broken power-law
luminosity function for the non-evolving luminosity $L_0$,
\begin{equation}
\psi(L_0) \propto
\begin{cases}
L_0^{-\alpha_1}, & L_0 < L_b, \\
L_0^{-\alpha_2}, & L_0 > L_b,
\end{cases}
\end{equation}
with $\alpha_1 = 0.17$, $\alpha_2 = 1.33$, and
$L_b = 1.33 \times 10^{41} \, \mathrm{erg\,s^{-1}}$. In the Monte Carlo
implementation, luminosities are drawn analytically by inverting the cumulative
distribution, assigning a fraction of $60\%$ of the events to the low-luminosity
branch. The intrinsic luminosity is then allowed to evolve with redshift as
\begin{equation}
L(z) = L_0 (1+z)^k,
\end{equation}
where $k = 6.66$ quantifies the luminosity evolution.

\begin{figure}[h]
    \centering
    \includegraphics[width=0.7\linewidth]{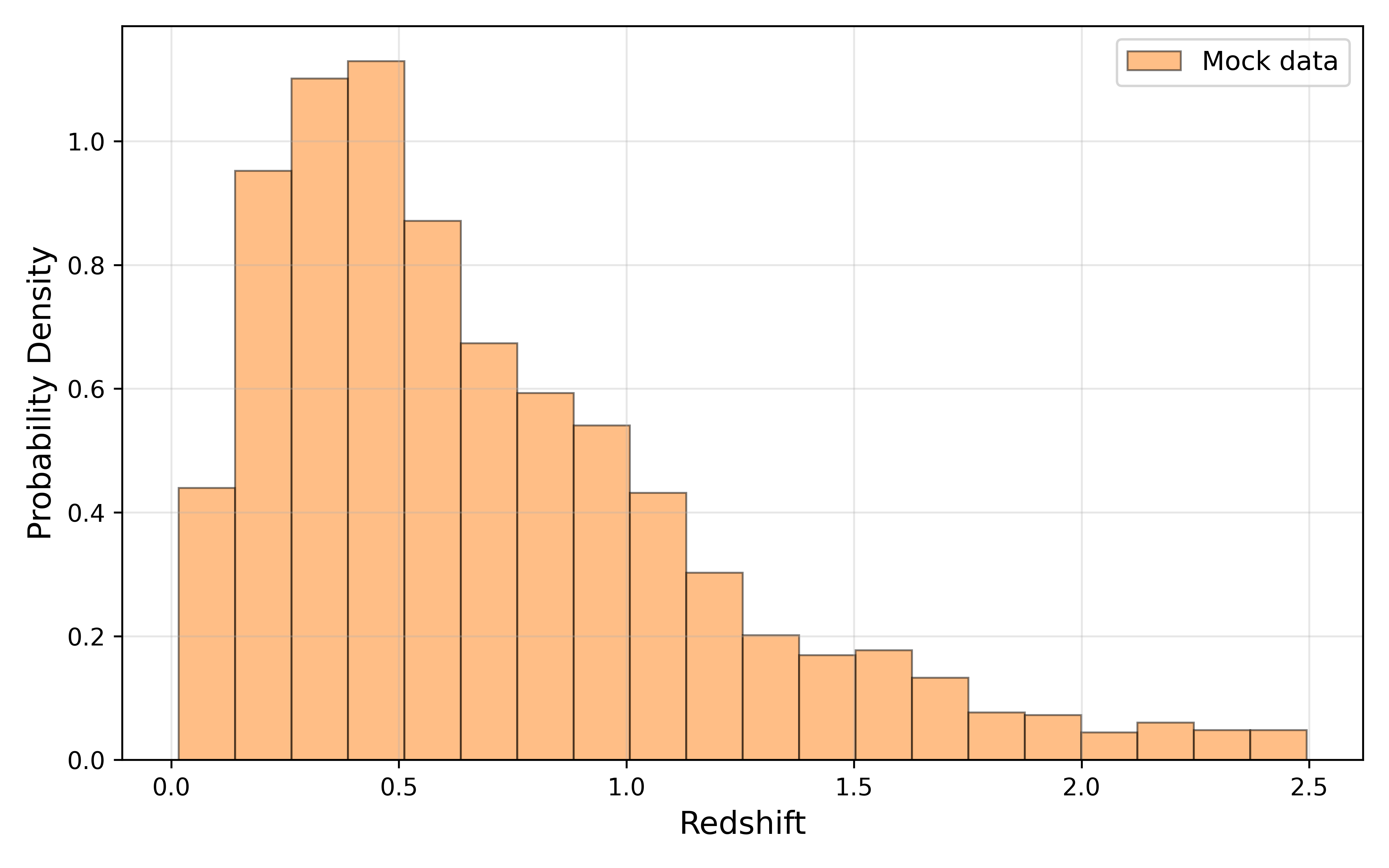}
    \caption{Redshift distribution of the detected mock FRB catalog obtained
    after applying the CHIME-like flux selection.}
    \label{fig_redshift}
\end{figure}

The observed flux density is computed as
\begin{equation}
F_\nu =
\frac{L(z)}
{4\pi D_L^2(z) \, \Delta\nu},
\end{equation}
and converted to Jy before applying the selection cut. We impose a detection
threshold
\begin{equation}
F_\nu > F_{\rm lim},
\end{equation}
with $F_{\rm lim} = 0.2~{\rm Jy}$ and
$\Delta\nu=400~{\rm MHz}$, mimicking a simple CHIME-like selection effect. The
mock catalog is constructed by iterating this procedure until 2000 detected FRB
events are obtained. Fig.~\ref{fig_redshift} shows the resulting redshift
distribution under these assumptions.

We then assign dispersion measures to the simulated events. The IGM contribution
is generated by introducing a stochastic fluctuation $\Delta$ around the mean
IGM dispersion measure. Specifically, we assume the quasi-Gaussian cosmic PDF
for $\Delta$ given by Eq.~\eqref{pdf_distri}, with
$\alpha=\beta=3$. The normalization parameter $C_0$ is fixed numerically by
requiring $\langle \Delta \rangle =1$. For each redshift, the scatter is taken
as
\begin{equation}
\sigma_{\rm IGM}=Fz^{-1/2},
\end{equation}
with $F=0.32$, following Ref.~\cite{Macquart:2020lln}. The IGM dispersion
measure is then drawn as
\begin{equation}
{\rm DM}_{\rm IGM}(z)=
\Delta\,\langle{\rm DM}_{\rm IGM}(z)\rangle,
\end{equation}
where $\langle{\rm DM}_{\rm IGM}(z)\rangle$ is computed from the fiducial
cosmological model.

The host-galaxy contribution is modeled as a lognormal distribution, following
Eq.~\eqref{p_host}. In the mock generation, we adopt $\mu=5.14$ and $\sigma_{\rm host}=0.83$. Finally, the total extragalactic dispersion measure is computed as
\begin{equation}
{\rm DM}_{\rm ext}
=
{\rm DM}_{\rm IGM}
+
\frac{{\rm DM}_{\rm host}}{1+z},
\end{equation}
as in Eq.~\eqref{DM_ext_th}, yielding the final mock catalog of 2000 detected
FRBs. 



\bibliographystyle{JHEP}
\bibliography{biblio.bib}






\end{document}